\documentclass[prb,superscriptaddress,twocolumn,showpacs,preprintnumbers,
amsmath,amssymb]{revtex4}

\usepackage{graphicx} 
\usepackage{dcolumn} 
\usepackage{bm}
\usepackage{txfonts}
\usepackage{mathrsfs}
\usepackage{comment}

\usepackage[]{psfrag}

\psfrag{t_rho}{$\tilde{\rho}$}
\psfrag{T_c}{$T_{c}$}
\psfrag{d_del}{$\delta-\delta_{\text{c}}$}
\psfrag{pp_u}{$\tilde{u}$}

\psfrag{tyldu}{$\tilde{u}$}
\psfrag{tyldel}{$\tilde{\delta}$}
\psfrag{delta}{$\delta$}
\psfrag{logdelt}{$\log (\delta-\delta_c)$}
\psfrag{logphi}{$\log (\phi_0)$}
\psfrag{dgdc}{$\frac{\delta_G-\delta_c}{\delta_c}$}
\psfrag{Temp}{$T$}
\psfrag{eta}{$\eta$}

\psfrag{tyldu}{$\tilde{u}$}
\psfrag{deltal}{$\delta$}
\psfrag{logdelt}{$\log (\delta-\delta_c)$}
\psfrag{logphi}{$\log (\phi_0)$}
\psfrag{dgdc}{$\frac{\delta_G-\delta_c}{\delta_c}$}
\psfrag{Temp}{$T$}
\psfrag{etaom}{$\eta_\omega$}
\psfrag{delta}{$\delta$}
\psfrag{phi}{$\phi_{0}$}
\psfrag{beta=0.13}{$\beta=0.13$}
\psfrag{beta=0.29}{$\beta=0.29$}
\psfrag{beta=0.50}{$\beta=0.50$}
\psfrag{log(phi)}{$\log\left[\phi_{0}\right]$}
\psfrag{log(delta-delta0)}{$\log\left[\delta-\delta_{c}\right]$}

\def\b0{{\mathbf 0}}

\def\Lam{\Lambda}

\begin{document}


\title{Phase boundary and finite temperature crossovers of the quantum Ising model in two dimensions}

\author{P. Strack}
\email{p.strack@fkf.mpg.de}
\affiliation{Max-Planck-Institute for Solid State Research,
Heisenbergstr.~1, D-70569 Stuttgart, Germany}

\author{P. Jakubczyk}
\affiliation{Max-Planck-Institute for Solid State Research,
Heisenbergstr.~1, D-70569 Stuttgart, Germany}
\affiliation{Institute for Theoretical Physics, Warsaw University, 
Ho\.za 69, 00-681 Warsaw, Poland}

\date{\today}

\begin{abstract}

We revisit the two-dimensional quantum Ising model by computing renormalization group flows close to its quantum critical point. The low but finite temperature regime 
in the vicinity of the quantum critical point 
is squashed between two distinct non-Gaussian fixed points: the classical fixed point dominated by thermal fluctuations and the quantum critical fixed point dominated by zero-point quantum fluctuations. Truncating an exact flow equation for the effective action we derive a set of renormalization group equations and analyze how the interplay of quantum and thermal fluctuations, both non-Gaussian in nature, influences the shape of the phase boundary and the region in the phase diagram where critical fluctuations occur. 
The solution of the flow equations makes this interplay transparent: we detect finite temperature crossovers by computing critical exponents and we confirm that the power law describing the finite temperature phase boundary as a function of control parameter is given by the correlation length exponent at zero temperature as 
predicted in an $\epsilon$-expansion with $\epsilon=1$ by Sachdev, Phys. Rev. B {\bf 55}, 142 (1997). 


\end{abstract}
\pacs{05.10.Cc, 73.43.Nq, 71.27.+a}

\maketitle

%
%
%
The quantum Ising model serves as a prime textbook example to illustrate fundamental 
aspects of quantum phase transitions.\cite{sondhi97,sachdev99,stewart01,belitz_review05, loehneysen_review07} The quantum Ising Hamiltonian has the form,
\begin{eqnarray}
H_{\text{QI}}=-J\sum_{\langle ij\rangle}\sigma^{z}_{i}\sigma^{z}_{j}-
h\sum_{i}\sigma_{i}^{x}\;,
\label{eq:H_QI}
\end{eqnarray}
where $J$ is a ferromagnetic exchange coupling, the sum $\langle ij\rangle$ runs 
over pairs of nearest neighbor sites,  and the quantum degrees of freedom are 
represented by the operators $\sigma_{i}^{z,x}$ which reside on a site $i$ of a hypercubic lattice in $d$ dimensions and reduce to the Pauli matrices in the basis where $\sigma^{z}$ is diagonal.\cite{sachdev99} The parameter $h$ is the external transverse magnetic field which induces quantum-mechanical tunneling events that flip the orientation of the Ising spins. 
The relevant parameter of Eq. (\ref{eq:H_QI}) is the ratio 
$\hat{\delta}\sim J/h$. For large $\hat{\delta}$ 
the ground state is ferromagnetically ordered and spontaneously breaks the discrete $Z_{2}$ Ising symmetry while for smaller $\hat{\delta}$ the spins in the ground state remain 
disordered. The two phases are separated by a second order quantum phase transition at a critical $\hat{\delta}_{c}$. At finite temperature the formation of spin order is hindered and the $\hat{\delta}$ at which the order sets in is increased leading to a 
line of second order phase transitions $T_{c}\left(\hat{\delta}\right)$ that terminates at the quantum critical point (QCP) $T_{c}\left(\hat{\delta}_{c}\right)=0$. 
Since the phase diagram of the quantum Ising model exhibits many generic features of physical systems in vicinity of their QCPs, it is important to understand it in detail.

Various finite temperature properties of compounds modelled by 
the quantum Ising model were 
measured experimentally in three dimensions.\cite{bitko96} 
Theoretically, the corresponding phase diagrams were 
investigated by Sachdev within analytical approaches.\cite{sachdev97,sachdev99}
These rely on the effective continuum field theory to which an expansion 
around the upper critical dimension is applied. In two dimensions, the quantum Ising model describes a strongly coupled lattice system. Its ground state was recently analyzed numerically with new algorithms.\cite{jordan08, cincio08}
The perturbative renormalization group (RG) approach by Hertz \cite{hertz76} 
and Millis \cite{millis93} does not cover this case as the QCP is associated with a non-Gaussian fixed point therefore invalidating the -- in other cases successful -- expansion around a Gaussian fixed point. \cite{millis93}

In this note, we extend our recent RG approach \cite{jakubczyk08} to QCPs associated with non-Gaussian fixed points. We present an analysis addressing the quantum Ising model in two spatial dimensions near the QCP with flow equations derived within the functional renormalization group framework.\cite{berges_review02} This set of coupled differential equations is valid at zero and finite temperature and is derived from a truncation of the exact functional flow equation for the scale-dependent effective action $\Gamma[\phi]$, with $\phi$ a scalar-valued bosonic field 
obtained from coarse-graining Ising spins over a neighborhood of their lattice sites. The solution of this flow equation as a function of the continuous cutoff scale $\Lambda$ yields the renormalized effective action from which all physical properties 
can be extracted. Already in simple truncations, this framework yields the critical properties of $O(N)$-symmetric field theories below the upper critical dimension,\cite{berges_review02} including the Ising case $O(1)$.\cite{ballhausen04} 

The scale-dependent action parametrizing the continuum field theory for the low energy physics of the quantum Ising model,\cite{sachdev97,sachdev99} which we will apply in this note is given by 
\begin{eqnarray}
 \Gamma_{\text{QI}}[\phi] =
 \frac{T}{2} \sum_{\omega_n} \int\frac{d^dp}{(2\pi)^d} \,
 \phi_p Z\left(\omega^{2}_{n} + \,\mathbf{p}^{2} \right) \phi_{-p} + U[\phi] \; .
 \label{eq:quantum_ising}
\end{eqnarray}
Here $p=\left(\omega_{n},\mathbf{p}\right)$ and $\omega_{n}=2\pi n T$ with $n$ integer are bosonic Matsubara frequencies, $Z$ is a $\Lambda$-dependent renormalization factor multiplying the momentum dependence of the propagator, and $U[\phi]$ is the effective potential specified below. For simplicity, the renormalization factors corresponding to the $\mathbf{p}^2$ and $\omega_{n}^{2}$ terms are taken to be equal here. 
As we checked by an explicit calculation, considering two different $Z$-factors for these two terms has no impact on the results of this paper. The action is regularized in the ultraviolet by restricting momenta to 
$|\mathbf{p}|<\Lambda_{\text{UV}}$. 
By virtue of the quadratic frequency dependence, the dynamical exponent $z$ is equal to unity. The effective dimensionality at zero temperature $\mathcal{D}=d+z=3$ is below the upper critical dimension $\mathcal{D}^{+}=4$. We approach the phase boundary and the QCP from the symmetry-broken region in the phase diagram and we therefore assume a potential $U[\phi]$ with a minimum at a
non-zero order parameter $\phi_0$:
\begin{eqnarray}
 U[\phi] &=& \frac{u}{4!} \int_0^{1/T} d\tau \int d^d x
 \left( \phi^{2} - \phi_{0}^{2} \right)^{2} \nonumber\\
 &=& \int_0^{1/T} \!\!\! d\tau \int \! d^d x
 \left[u\,\frac{\phi'^{4}}{4!}+\sqrt{3 \, u\, \delta} \,
 \frac{\phi'^{3}}{3!} + \delta \, \frac{\phi'^{2}}{2!}\right] \, ,
\label{eq:ef_potential}
\end{eqnarray}
where $\phi$ and $\phi'$ are functions of $x$ and $\tau$
with $\phi= \phi_{0}+\phi'$.
The parameter
$\delta = \frac{u\,\phi_{0}^{2}}{3}$ is related to the ratio of transverse field to exchange coupling $\hat{\delta}\sim J/h$ in the original Hamiltonian, Eq. (\ref{eq:H_QI}), and 
controls the distance from criticality. The three-point vertex $\sqrt{3\, u\, \delta}$ 
generates an anomalous dimension of the order parameter field
already at one-loop level. 

%
The flow equations are obtained along the lines of Ref.~\onlinecite{jakubczyk08}. 
The recipe is the following: after endowing the propagator with a suitably chosen 
cutoff function (also given in Ref.~\onlinecite{jakubczyk08}) that implements the $\Lambda$-dependence and 
regularizes the infrared singularity of the massless propagator at criticality, one executes a cutoff-derivative 
on the analytic expressions corresponding to all one-loop one-particle irreducible 
Feynman diagrams for the parameters $u$, $\phi_{0}$. The flow of $Z$ is obtained by taking second derivative of the equation describing the flow of the propagator with respect to momentum. After utilization of the variables
\begin{eqnarray}
\tilde{\rho}=\frac{\phi_{0}^{2}\,Z d}{2K_{d}T\Lam^{d-2}}\,\,,\hspace*{10mm} 
\tilde{u}=\frac{u\,2 K_{d}T}{d Z^{2}\Lam^{4-d}}\;, 
\end{eqnarray}
where $K_{d}$ is defined via $\int \frac{d^{d}k}{\left(2\pi\right)^{d}
}=K_{d}\int d|\mathbf{k}|\,|\mathbf{k}|^{d-1}$, as well as the anomalous dimension $\eta$ and rescaled temperature $\tilde{T}$
\begin{eqnarray}
\eta = - \frac{d\log Z}{d\log\Lambda}\,\,,\hspace*{10mm}
\tilde{T}=\frac{2\pi T}{\Lambda}\;,
\label{eq:eff_temp_QI}
\end{eqnarray}
we can write the flow equations as 
\begin{eqnarray}
\frac{d\tilde{u}}{d\log \Lambda}&=&\left(d-4+2\eta\right)\tilde{u}
+\nonumber\\
&&
3\tilde{u}^{2}
\left[
\frac{1}
{\left(1+\frac{2\tilde{u}\tilde{\rho}}{3}\right)^{3}}
+
2\sum_{n=1}^{\infty}
\frac{1}
{\left(\left(n \tilde{T}\right)^{2} + 1+ \frac{2\tilde{u}\tilde{\rho}}{3}\right)^{3}}
\right]\nonumber\\
\frac{d\tilde{\rho}}{d\log \Lambda}&=&\left(2-d-\eta\right)\tilde{\rho}
+\nonumber\\
&&\frac{3}{2}
\left[
\frac{1}
{\left(1+\frac{2\tilde{u}\tilde{\rho}}{3}\right)^{2}}
+
2\sum_{n=1}^{\infty}
\frac{1}
{\left(\left(n \tilde{T}\right)^{2} + 1+ \frac{2\tilde{u}\tilde{\rho}}{3}\right)^{2}}
\right].
\label{eq:finiteT_u_rho_QI}
\end{eqnarray}
The anomalous dimension is determined by
\begin{eqnarray}
\eta&=&2\tilde{u}^{2}\tilde{\rho}
\Bigg[
\frac{1}
{\left(1+\frac{2\tilde{u}\tilde{\rho}}{3}\right)^{4}}
-
\frac{2}
{\left(d+2\right)\left(1+\frac{2\tilde{u}\tilde{\rho}}{3}\right)^{5}}
+\nonumber\\
&&2\sum_{n=1}^{\infty}
\frac{1}
{\left(\left(n \tilde{T}\right)^{2}+1+\frac{2\tilde{u}\tilde{\rho}}{3}\right)^{4}}
-
\frac{2}
{\left(d+2\right)\left(\left(n \tilde{T}\right)^{2}
+1+\frac{2\tilde{u}\tilde{\rho}}{3}\right)^{5}}
\Bigg]
\;.\nonumber\\
\label{eq:finiteT_eta_QI}
\end{eqnarray}
The Matsubara sums can be performed 
analytically yielding hyperbolic functions but the expressions 
do not deliver any additional insights here. 
There are three distinct contributions in the flow equations 
Eq. (\ref{eq:finiteT_u_rho_QI}): the Gaussian terms linear in 
$\tilde{\rho}$, $\tilde{u}$, the \emph{classical} terms corresponding to the zeroth Matsubara frequency, and the $\emph{quantum}$ terms summing over the non-zero Matsubara frequencies that become crucial at low temperatures. The zero temperature limit of 
Eq. (\ref{eq:finiteT_u_rho_QI}) results in modified variables 
$\tilde{\rho}$ and $\tilde{u}$, which then do not depend on $T$ and are 
rescaled by different powers of $\Lambda$, but is fully accessible within our framework.\cite{jakubczyk08,phd09}  A strength of the present approach 
lies in the ability to integrate out classical and quantum fluctuations -- both \emph{non-Gaussian} in nature, including the anomalous dimension $\eta$ of Eq. (\ref{eq:finiteT_eta_QI}).\cite{jakubczyk08}

\begin{figure}[t]
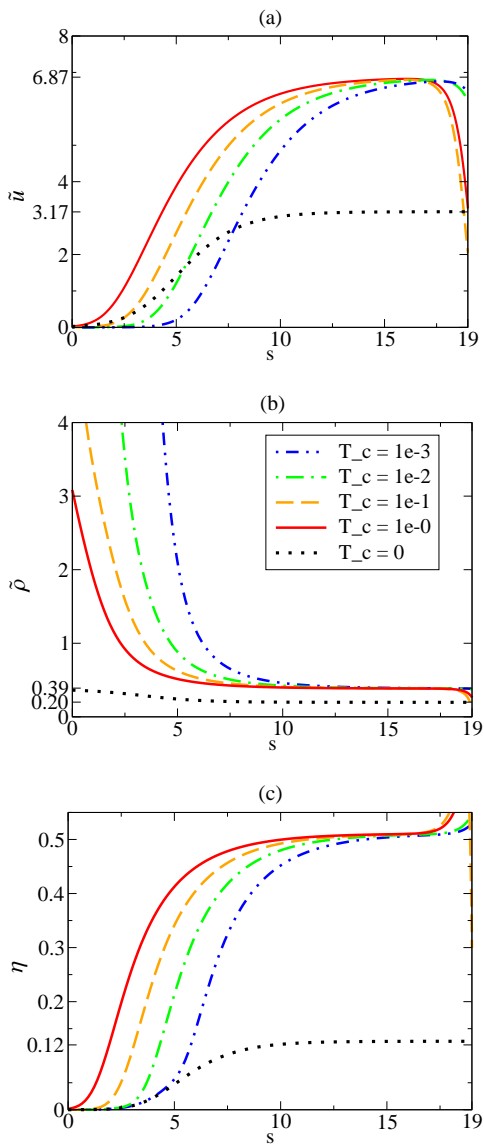

\includegraphics*[width=62mm]{u_flows_z_1.eps}\\[4mm]
\includegraphics*[width=63mm]{rho_flows_z_1.eps}\\[4mm]
\includegraphics*[width=63mm]{eta_flows_z_1.eps}
\caption{(Color online) Flows for the quantum Ising model in $d=2$ as a function of logarithmic cutoff-scale
$s=-\log\left[\Lam/\Lam_{\text{UV}}\right]$ for various temperatures and at $T=0$. We set $\Lambda_{\text{UV}}=1$.
The infrared (ultraviolet)
is to the right (left) of the graphs. The
values of the classical fixed point, attained by all finite-$T$ flows, and the 
quantum critical fixed point, attained by the zero-temperature flow, are marked on the vertical axis. 
(a): Flows of the quartic self-interaction $\tilde{u}$. 
(b): Corresponding flows of the rescaled minimum of the effective potential $\tilde{\rho}$. 
(c): Flows of the anomalous dimension $\eta$. }
\label{fig:u_rho_flows_z_1}
\end{figure}

The numerical solution of the coupled flow equations 
(\ref{eq:finiteT_u_rho_QI}, \ref{eq:finiteT_eta_QI}) at $T=0$ 
gives access 
to the critical behavior at the QCP while for $T>0$ 
the classical critical behavior along the phase boundary is obtained. As initial conditions for the upper cutoff, the interaction coupling, and the momentum renormalization factor we set $\Lambda_{\text{UV}}=1$, $u=1$, and $Z=1$. By choosing 
the initial condition of $\tilde{\rho}$ or $\delta$, we can tune to the critical state 
characterized by $\delta$ vanishing at the end of the flow, that is, for $\Lambda\rightarrow 0$.\cite{jakubczyk08} 
In Fig. \ref{fig:u_rho_flows_z_1}, the zero temperature flow at the QCP is juxtaposed with finite temperature 
flows at the phase boundary. Both, the finite-$T$ 
and the $T=0$ theory find a description in terms of two distinct non-Gaussian fixed points indicated by two distinct scaling plateaus with finite $\tilde{u}$ and $\eta$. The anomalous dimension directly at the QCP (black dots in Fig. \ref{fig:u_rho_flows_z_1} (c)) comes out as $\eta_{\text{QCP}}\approx0.1$ to be compared with the accurate value of the classical Ising universality class in 3D:\cite{pelisetto02} $\eta\approx0.04$. The correlation length exponent at $T=0$ comes out as $\nu=0.6$ in our calculation -- coincidentally close to the currently accepted value $\nu=0.63$.\cite{pelisetto02} We calculate $\nu$ by first computing the susceptibility 
exponent $\gamma$ and then using the scaling relation $\gamma=\nu\left(2-\eta\right)$. 
\cite{goldenfeld92} The susceptibility corresponds to the renormalized value of 
$\delta^{-1}_{\Lambda\rightarrow0}$ at the end of the flow. The exponent 
$\gamma$ is obtained as the slope of the graph of $\delta^{-1}_{\Lambda\rightarrow0}$ versus the distance from the QCP $\left(\delta-\delta_{c}\right)_{\Lambda=\Lambda_{\text{UV}}}$ in double logarithmic coordinates. 

At the finite temperature fixed point, $\eta\approx0.5$, to be compared with the exact value from the Onsager solution of the classical Ising model in two dimension $\eta=1/4$.\cite{goldenfeld92} More elaborate truncations lead to improved accuracy in the critical exponents.\cite{canet03,ballhausen04,benitez09}

From the finite-temperature flows in Fig. \ref{fig:u_rho_flows_z_1} (a), we can deduce the Ginzburg-scale, where $\tilde{u}$ starts to become sizable, to vary with temperature as
$\Lambda_{G}\propto T_{c}^{1/\left(4-d\right)}$,
with $d=2$, thereby fitting the formula valid for $d+z>4$.\cite{jakubczyk08} 
Juxtaposing the Ginzburg scale with the quantum-to-classical crossover scale which follows from the definition 
of $\tilde{T}$ in Eq. (\ref{eq:eff_temp_QI}): $\Lam_{cl}\sim T^{1/z}$ with $z=1$, we obtain
\begin{eqnarray}
\Lambda_{G}\sim T^{1/2}>\Lambda_{cl}\sim T\;,
\end{eqnarray}
which indicates that non-Gaussian fluctuations become important at energy scales above the quantum-to-classical crossover. For the cases where the QCP is described by a Gaussian fixed point, \cite{jakubczyk08} this relation is inverted
$\Lambda_{G}<\Lambda_{cl}$, while $\Lambda_{G}\approx\Lambda_{cl}$ for $d=2$, $z=2$. 

Another important difference between the QCP being Gaussian or non-Gaussian is that in the latter case $\Lambda_{G}$ does not vanish as $T\rightarrow0$. In the $(T,\tilde{u})$-plane there is a jump from the $2D$-Ising 
fixed point at finite temperature to the $3D$-Ising fixed point at $T=0$ with a finite $\Lambda^{T=0}_{G}$ as shown by the zero-temperature flow in Fig. \ref{fig:u_rho_flows_z_1} (a).

\begin{figure}[t]
\vspace*{1.5mm}
\includegraphics*[width=68mm,angle=0]{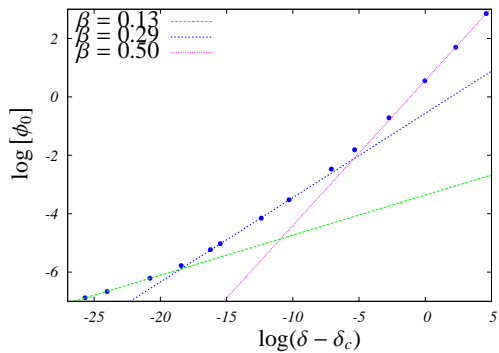}
\caption{(Color online) Emergence of three different regimes in the phase diagram as illustrated with three different values of the order parameter exponent $\beta$ upon 
approaching the QCP at very low temperatures. Here, we set $T=1.7\times10^{-5}$.}
\label{fig:beta_crossover}
\end{figure}

\begin{figure}[b]
\hspace*{-2mm}
\includegraphics*[width=60mm,angle=0]{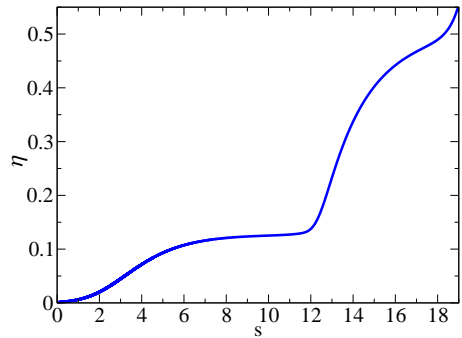}
\caption{(Color online) Crossover behavior of the anomalous dimension 
as a function of scale $s=-\log\left[\Lam/\Lam_{\text{UV}}\right]$ for 
$T=1\times10^{-6}$. Upon increasing $T$, this curve 
will continuously deform toward the shape of the blue dashed-dotted line 
in Fig. \ref{fig:u_rho_flows_z_1} (c).}
\label{fig:eta_crossover}
\end{figure}

The phase diagram in the symmetry-broken phase 
at low but finite temperatures can be portioned into three regimes, characterized by exponents belonging to different universality classes as shown 
for the order parameter exponent $\beta$ in Fig. \ref{fig:beta_crossover} and for the anomalous dimension $\eta$ as a function of scale in Fig. \ref{fig:eta_crossover}. 
The regimes and their 
relative size in the phase diagram can be detected by computing $\beta$ when approaching the phase boundary from the symmetry-broken phase: $\phi_{0}\sim\sqrt{\rho}\sim\left(\delta-\delta_{c}\right)^{\beta}$.
The results in double logarithmic coordinates, so that the slope corresponds to the exponent $\beta$, are exhibited in Fig. \ref{fig:beta_crossover}.
In the immediate vicinity of the phase boundary, $\beta\approx0.13$, which comes out close to the exact value $\beta=1/8=0.125$ 
from the Onsager solution 
of the classical $2D$-Ising model.\cite{goldenfeld92}
This reflects the fact that at any finite, even if small, temperature the asymptotic properties of the system are determined by strong 
classical, non-Gaussian fluctuations. This asymptotic behavior 
does not survive the limit $T\rightarrow0$, as the classical fluctuations are replaced by 
zero-point quantum fluctuations. Further away from the phase boundary, 
in the center of Fig. \ref{fig:beta_crossover}, $\beta\approx0.29$, which comes 
out close to the value of the $3D$-Ising universality class $\beta=0.33$.\cite{pelisetto02} This regime persists at zero temperature 
and reflects the non-Gaussian character of the QCP. Further away from the phase boundary, to the right of Fig. \ref{fig:beta_crossover}, mean-field behavior sets in 
with $\beta\approx0.5$.

The same crossover between the three regimes manifests itself also in the 
scaling behavior of the propagator represented by the anomalous dimension in Fig. 
\ref{fig:eta_crossover}. In the high energy regime of the flow ($0\leq s\lesssim2$), 
$\eta$ is close to zero reflecting mean-field behavior. At lower energies, 
between $s\sim5$ and $s\sim12$, the flow is governed by the quantum critical fixed 
point which belongs to the 3d-Ising universality class with a scaling plateau at $\eta\sim0.1$. Asymptotically in the infrared 
($15\lesssim s \lesssim18$), classical scaling of the 2d-Ising universality 
class sets in with $\eta$ being attracted toward $\eta\sim0.5$. Upon 
increasing $T$, the size of the classical 2d-Ising plateau 
increases and the size of the quantum 3d-Ising plateau decreases. 

By identifying the finite temperature phase boundary with the critical coordinates in the $(\delta,T)$-plane where the non-Gaussian scaling plateaus in Fig. \ref{fig:u_rho_flows_z_1} occur,\cite{jakubczyk08} we have computed 
the critical temperature as a function of the control parameter and 
obtained the power law
\begin{eqnarray}
T_{c}\sim\left(\delta-\delta_{c}\right)^\nu\; 
\label{eq:shift}
\end{eqnarray}
with $\nu=0.6$ the correlation length exponent at zero temperature as determined above. Sachdev obtained the same result for $T_{c}$ expanding around the upper critical dimension in $\epsilon=3-d$ and then extrapolating the result to $d=2$. \cite{sachdev97} Although yielding the correct result, at least the formal justification of setting $\epsilon=1$ without capturing the anomalous dimension and the associated non-Gaussian behavior in the strong-coupling region of the phase diagram is not immediately clear to us.

\begin{figure}[t]
\vspace*{4mm}
\hspace*{2mm}
\includegraphics*[width=72mm,angle=0]{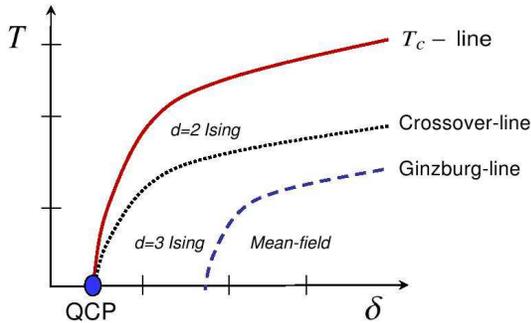}
\caption{(Color online) Schematic plot of the Ginzburg-line (dashed, blue), the crossover line (dotted, black) from $3D$-Ising to $2D$-Ising behavior, and the 
true $T_{c}$-line (straight, red) for the quantum Ising model in $d=2$. 
Units are arbitrary and will depend on the microscopic 
details of the specific system under investigation. 
The symmetry-broken phase is the area to the right of the $T_{c}$-line.}
\label{fig:ginz_vs_Tc}
\end{figure}

In Fig. \ref{fig:ginz_vs_Tc}, we schematically 
plot the $T_{c}$-line, the crossover line 
from $3D$-Ising to $2D$-Ising behavior, and the Ginzburg temperature (at which the interaction $\tilde{u}$ becomes sizable). 
Relying on the Ginzburg-line as a proxy for the phase boundary gives a different location of the QCP. 

In conclusion, we have extended our recent RG framework for quantum-critical systems with discrete symmetry-breaking \cite{jakubczyk08} to 
systems where the QCP is associated with a non-Gaussian fixed point. As a stresstest, 
we performed an RG analysis of the low but finite temperature regime of the two-dimensional quantum Ising model where the system is squashed between the strong-coupling fixed points of (i) the QCP and (ii) the finite temperature phase boundary.

Interesting avenues for future investigation include the as yet unresolved 
interplay of quantum and thermal fluctuations in the vicinity of a QCP where 
the finite temperature phase transitions are of the Kosterlitz-Thouless type. 

We thank S. Sachdev for a useful correspondence and 
A. Muramatsu for providing references. 
W. Metzner is acknowledged for valuable discussions and critically reading the manuscript. 
PJ acknowledges support from the German Science Foundation through 
the research group FOR 723.


\end{document}